\documentstyle[twocolumn,aps]{revtex}
\begin{document}

%
% Title Page
%

\title{ Where are we in the theory of High Temperature 
Superconductors\footnote{Talk given at the K S Krishnan
Birth Centenary, Conference on Condensed Matter Physics, 
Department of Physics, University of Allahabad,
Allahabad, India; 4-7 Dec 1998 }?}

\author{G. Baskaran\cite{email}}

\address{ The Institute of Mathematical Sciences,
                      Madras 600 113, India}

\maketitle

\begin{abstract}

In this talk I will briefly review our present theoretical 
understanding of some of the important issues in the high Tc
cuprates.  In view of its success, at a qualitative level and 
some times quantitative level, the theory initiated by Anderson and  
developed further by him and collaborators will be discussed. 
Many issues that still challenges us will be pointed out.  

\end{abstract}

\section {Introduction}

I feel honored to speak at this conference which commemorates 
the Birth Centenary of an eminent son of India, Sir K S Krishnan,
the co-discoverer of Raman effect,  and the Platinum Jubilee year
of this Physics Department that has nurtured many excellent physicists 
over the years.  Krishnan would have enjoyed seeing  
the development in the field of high Tc superconductivity, 
in which a two dimensional metallic character and quantum magnetism 
play fundamental role.  Krishnan\cite{ksk}, in late 30's,  pioneered 
the study of anisotropic transport and magnetic properties of graphite, 
an excellent example of two dimensional metal.  He had deep 
insights in the magnetism of transition metal and rare earth ions 
through his innovative susceptibility measurements of 
families of magnetic systems such as $Cu SO_4.5H_2O$, that are actually 
Mott insulators in the current parlance.

Nearly 12 years ago Bednorz and Muller\cite{bednorz} discovered 
superconductivity in Ba doped LSCO and broke the barrier of the 
record Tc of 23 K exhibited by the A-15 family member $Nb_3 Ge$.  
Soon many cuprates 
were synthesized and now the maximum Tc is nearly 160 K in some Tl/Hg
based cuprate under pressure.  From experiment point of view the 
quality of single crystals have improved considerably over years 
and we have a good set of several reproducible experimental results 
that begs quantitative explanation.

From theory point of view the Resonating Valence Bond (RVB) theory of 
Anderson\cite{pwascience} that had a lead right from the beginning in view 
of its strong foundation on the available body of experimental facts, has made 
significant progress, considering the nature of the hard quantum many body 
problem that cuprates posed. This fertile theory, by its fairly penetrating
character  also initiated a resurgence in the theory of strongly 
correlated systems and quantum magnetism.  This theory has remained the 
leading guidance in the sense of providing the right directions 
and emphasizing the crucial aspects of the problem, albeit with occasional
changes compelled by new experimental results.   
  
In this talk I will enumerate some of the important issues and point 
out our current understanding from the point of RVB theory.  This talk
will be sketchy and some details and many helpful hints for making 
further progress can be found in Anderson's book\cite{pwabook}.

\section{Anderson's Original Proposal and initial progress}

Anderson's original proposal\cite{pwascience}  
presented at the Bangalore conference
in January 87 identified the relevant interactions and presented
it in a succinct form as a one band large U Hubbard model or the
equivalent t-J model.  The insulating parent compound LSCO was 
suggested to be a 2d Mott insulator in a disordered spin liquid or
RVB state.  The resonating spin singlets are neutral in the 
insulating state - they do not transport charges at low energies.
On doping they start transporting charges leading to 
superconductivity.  Anderson also suspected the presence of 
neutral spin half excitations(which was later named spinons\cite{ABZH})
with their own pseudo fermi surface.  RVB mean field theory\cite{BZA} 
that brought out the neutral spinons and their pseudo fermi surface
in the insulating state and superconductivity in the doped case were
discussed by Anderson and collaborators.  Affleck, Martson and
Kotliar\cite{flux} brought out an energetically better mean field state 
namely the d-RVB or the flux state.  Inspired by Anderson's suggestions
Kivelson and collaborators\cite{KRS} discussed short range RVB 
in some detail focusing on spinon and holon excitations.  
Slave boson theories and gauge theories\cite{gauge} followed suit and 
there were intense activities and speculations, including the possible 
parity violating superconducting states\cite{anyon} with connection 
to Laughlins quantum Hall state. 

Looking back, the proposals of Anderson, with its emphasize
on strong correlation, the ensuing non fermi liquid states, spin
charge decoupling, spinon fermi surface,  has remained robust 
and has given us a good way to think about this complex problem.  
In particular the ARPES, neutron scattering, NMR relaxation and 
transport properties can be qualitatively understood from the point 
of view of the above proposal.  However, quantitative understanding 
is yet to be achieved.

The precise mechanism of superconductivity, particularly in the 
single layer materials, is some thing that has
eluded a sharper theoretical understanding so far, even though it was
one of the first issues that caught the attention of the condensed
matter community.  Very fundamental and new ideas have however emerged
through the notion of inter layer pair tunneling, which we will
discuss at the end. The electron kinetic energy gain as the origin 
of the superconducting condensation energy is also a novel aspect 
of the present system.

\section{2d Quantum Antiferromagnet and under doped regime}

A good understanding of the Mott insulator should help one to understand
the doped Mott insulator better.  In this insulating state, kinetic or 
super exchange dominates and only spin degrees of freedom governed
by Heisenberg Hamiltonian are present at low energies.
An emerging local gauge symmetry in the insulating state was found and 
formalized by Anderson and the present author\cite{gauge} as a gauge
theory.  It was later discovered \cite{wen} that this gauge field 
captures the physics of chiral fluctuations among the interacting spins
in the low doped regime. The d-RVB state or the Affleck-Marston-Kotliar
phase\cite{flux} can be thought of as a uniform RVB state\cite{BZA} 
in which $\pi$ fluxes are condensed at low temperatures.  

There are good theoretical indications 
that the 2d Heisenberg model has a long range antiferromagnetic order.
Several static and quasi static phenomenon are well explained by the
spin wave theory inspired non linear sigma model analysis\cite{chak}.
However, the dominant correlations in the ground state is that of
a d-RVB state; as suggested by Hsu\cite{hsu} it is meaningful to think 
of the ordered state as a spinon density wave in a spin liquid state.
The antiferromagnetic order is fragile and disappears at  about 
$ 1.5 \% $ of doping.  Recent ARPES study\cite{shen} in under doped and 
insulating layered cuprates point out that the d-RVB with its 
massless Dirac like 
spinon spectrum is a good reference state to describe the Mott 
insulating and the under doped state. This questions the real relevance
of the non linear sigma model in its present form, in the doped quantum 
melted region.

The physics in the under doped regime is complicated by disorder,
long range coulomb interaction, charge localization and micro
phase separation effects.  This is the origin of the stripe phase.   
Some theoretical work is going on in this regime\cite{emery}.
A recent work by Fisher\cite{mathew} and collaborators, addresses the 
issue of how an insulator to superconductor transition takes place in
the ground state by their theory of nodal fermi liquids.  Their 
reference state is a d-wave superconductor, where quantum fluctuations
induced by coulomb correlations drive a metal(or insulator) superconductor 
transition.  This theory captures some of the physics of the 
t-J model and over emphasizes pair fluctuation of charges.  There are 
also some fundamental questions whether we can have a boson metal ground
state for low doping\cite{seb}. 

\section{`Solving' the t-J model} 

Several experimental results indicated the validity of the 
t-J model for the conducting cuprates\cite{pwabook}.  From theory
point of view the derivation of the t-J model has been shown rather 
satisfactorily through sharpening of the Zhang-Rice singlet arguments 
through detailed cluster calculations\cite{schlutter}.  
The t-J model however remains unsolved in a satisfactory fashion,
in view of the on site constraint involved.  {\bf In the absence of
an exact solution or a good many body theory, a natural way
to solve the t-J model is to look for the next level of effective
theories, guided by experiments and some times theoretical arguments,  
that will point to the correct final solution }.

In constructing the next level of effective theories there has 
been considerable work using slave boson mean field theories and
the related gauge field theories\cite{gauge} .  This approach, 
though presents itself with many new possibilities that often 
agree with the experimental trends, involves uncontrolled  
approximations and is far from satisfactory as a quantitatively 
correct many body theory. These theories are in a sense reaction 
against the conventional restrictive fermi liquid type of perturbative 
many body theories that fail in these class of correlated conductors.  

In the numerical front there has been many efforts by several
groups \cite{elbio}.  But they have not been very helpful in 
our understanding the rich low energy physics 
offered by the t-J model - they most often capture some
high energy features and face serious difficulties when it 
comes to low energy physics.

In the analytical front there is hope however, in the sense 
explained below, arising  
from Anderson's proposal \cite{pwabook} of failure of fermi liquid 
theory in 2d Hubbard model for arbitrarily small repulsion and 
the associated notion of 2d tomographic Luttinger liquid.  This means
that $ U^* = 0 $ is an unstable fermi liquid fixed point as in 
the 1d Hubbard model;  and in principle the strong coupling 
non fermi liquid fixed point $U^* = \infty$ can be understood by 
a careful study of small U. 

Anderson, through a phase shift analysis of two particles on the
fermi surface in the $2k_F$ and singlet channel argues for the
presence of a singular Landau parameter that leads to a different
spin charge velocities on the fermi surface and anomalous exponent
for the electron propagator.  This point has remained 
controversial\cite{shank} and the present author has given some 
supporting arguments\cite{gb1} for Anderson's proposal.  The present 
author\cite{gbzero} has also brought out a related mechanism involving 
zero sound that could destabilize fermi liquid state in 2 and 3 dimensions.  

The presence of singular forward scattering, Anderson argues,
leads to the so called tomographic Luttinger liquid (TLL) state,
a non fermi liquid state exhibiting spin charge decoupling as well
as branch point singularities for electron propagator on the fermi 
surface.  The TLL theory of Anderson is essentially a Landau's
fermi liquid theory, but with a singular forward scattering.  
I view it as a natural generalization of Landau's fermi liquid 
theory in the following sense.  In Landau's fermi liquid theory
occupied low energy quasi particle states influence each other
pairwise only in a mean field fashion irrespective of their relative
momenta.  In the tomographic Luttinger liquid, those electrons that 
have vanishingly small  relative momenta (that is, belonging to a given 
tomograph) influence each other pairwise, in a non mean mean field
fashion, leading to a finite phase shift in the relative momentum
channel.  While electrons belonging to different tomographs do not influence
each other. This has profound consequences, as argued by Anderson.

In view of its simple form, TLL theory should lend itself 
to more detailed analysis\cite{akh} and comparison with experimental 
results, particularly of the normal state.  Through Andersons proposal,
which has a good phenomenological support, we have a possibility of 
analyzing the t-J model, a strong coupling limit,  through a Landau 
type of theory.  The parameters of this non fermi liquid theory can be 
determined from experiments.

This scenario should work well for the optimal doping regime and
beyond.  However, it becomes a difficult problem 
with new possibilities when we go to the under doped situation.  
The physics near this region has the spin gap phenomenon and it requires 
some fresh thinking, or to go back to old RVB ideas to make further 
progress.  As mentioned earlier, disorder and long range interactions
and the corresponding charge localization effects cloud the real
issue that we are after. 
 
\section{Normal state as a tomographic Luttinger liquid}

The normal state of cuprates is generally accepted as  
anomalous  and non fermi liquid like, thanks to a variety of 
experimental results -  frequency and temperature dependent 
conductivity, Hall effect, NMR relaxation, thermal conductivity, 
the non fermi liquid spectral functions seen in ARPES measurements 
and so on.  While the 
semi phenomenological theory of Anderson's tomographic Luttinger
liquid suggests anomalous exponents, a satisfactory derivation
of exponents and other details awaits further theoretical 
developments.   

At another level, the two scattering rates on the fermi surface, 
one corresponding to the longitudinal resistance that scales 
as $ T $ and the other rate measured from Hall angle\cite{ong}
 scaling as $ T^2 $ needs to be formalized.  Anderson's suggestions 
and heuristics remain as fundamental proposals that calls for a 
satisfactory derivation to make quantitative progress.   

The development of spin gap\cite{rice} at low temperatures becomes prominent
when we go to the under doped situation.  In terms of the old RVB
idea this has a natural explanation in terms of the development
of neutral spinon pair condensation. However, the importance of 
interlayer pair tunneling and interlayer super exchange in 
providing the spin gap phenomenon has also been suggested by 
Anderson and collaborators\cite{pwabook}. The real origin of 
the spin gap, identifying the correct in plane and inter plane
contributions, needs to be sharpened further, in view of the contrast
between the strong interlayer correlations in YBCO, compared to the 
equally high Tc compound Tl-2201 with weak interlayer interaction.

Charged stripes in the normal states are interesting phenomenon of 
localization of the heavy holes that also suppress superconductivity.  
It is becoming clear that they are not providing any obvious
mechanism for superconductivity.  However, there are some 
intriguing experimental suggestions at low doping of 
very low energy stripe and perhaps antiferromagnetic long 
range order in an incommensurate peak position in the 
superconducting states\cite{neutron}.  These are likely  
to be complications that are not very essential for our understanding 
of the underlying robust physics of superconductivity and anomalous 
normal state.  However, we need a good theoretical understanding of them.

\section{Confinement and inter layer pair tunneling}

The original proposal of Anderson relied on the neutral singlets
of the insulating RVB state getting charged on doping and leading
to superconductivity.  This suggestion was however challenged by
the inter layer regularity of the Tc in various cuprates.
So it was felt that perhaps the quantum fluctuations arising
from the strong correlations in a single plane are strong enough
to suppress superconductivity by enhanced gauge field or phase
fluctuations.  

Around the same time the notion of confinement was introduced by 
Anderson and Zou\cite{zou} by looking at the large 
anisotropy of the normal state 
resistivity - the ratio of the c-axis resistivity to ab-plane 
resistivity was too large compared to band effective mass anisotropy. 
It was then argued that the spin charge decoupling in the anomalous 
normal state strongly suppresses  the electron spectral weight close 
to the fermi surface leading to absence of coherent one electron 
transport between neighboring planes.  This is called confinement.
One electron kinetic energy between two conducting
planes is frustrated.  This frustration leading to incoherent transport
is in a way more subtle than the way one electron kinetic energy is 
frustrated in a Mott insulator.  There is excellent experimental 
support  for confinement phenomenon from optical sum rule
measurements\cite{bosov}.  

A pair of electrons on the fermi surface in a spin 
singlet state with zero center of mass momentum, however, retains
its identity without any quantum number fractionization.  Hence it  
can tunnel coherently between two planes.  The selective suppression 
of one electron coherent tunneling , and not of two electrons, 
is the origin of inter layer pair tunneling mechanism of 
superconductivity of Wheatley, Hsu and Anderson\cite{wha,pwabook}. 
The frustrated
one electron kinetic energy loss is gained by pair delocalization
between two planes.  

Anderson formalized the above through a BCS type of effective
Hamiltonian that expressed the major source of pair condensation
energy as a pair tunneling terms between two fermi liquid planes:
$$
H_{pair} = - \sum_k T_J(k) c^\dagger_{k\uparrow 1} c^\dagger_{-k\downarrow
1}
c^{}_{-k\downarrow 2} c^{}_{k\uparrow 2}
$$
and a small intra plane BCS type of scattering term non local in
k-space.
Notice that in the above term, the individual electron momentum is
conserved, making it a resonant tunneling between to planes.
It is this resonant character or local character in k-space that
leads to a large Tc proportional to the pair tunneling matrix 
element $T_J$  -  a non-BCS dependence of Tc on the pairing 
interactions.

At the level of models, formalizing Anderson's BCS type of 
formalism\cite{pwabook} is a very important issue.  Early
derivation by Muthukumar\cite{muthu} in terms of slave boson variables, 
can be modified to bring out the electron momentum conservation.  
However, a satisfactory derivation, that also addresses the issue of 
one electron incoherence\cite{pwabook,clark}  and two electron coherence 
between planes or chains is still needed.   The absence of bilayer 
splitting in the ARPES spectral functions
is a good indication for low energy one electron incoherence between
two layers.  

The interlayer coherence phenomenon and possibility of 
application to the closely related organic conductors has been
studied by some authors\cite{clark1}.  The present author has argued
that\cite{gb4} certain extreme sensitivity of the Tc of organic superconductors
to off chain or off plane disorder points towards the origin
of inter layer pairing mechanism of superconductivity in
these systems and an apparent violation of Anderson's theorem for 
c-axis disorder.

\section{ Is Inter Layer pair tunneling the only mechanism of
superconductivity in cuprates ?}

The pair tunneling effective Hamiltonian has been successfully
used by Anderson and collaborators\cite{pwabook} to understand 
the origin of large $T_c$ and also certain features of the gap function in
k-space.  In this context Anderson also proposed an important
test\cite{pwaplasmon} for the interlayer pair tunneling mechanism: 
since the
superconducting condensation energy arises primarily from pair tunneling 
between planes, the c-axis Josephson plasma energy should be
the same as the pair condensation energy.  This remarkable
prediction was verified\cite{iltexpt} for the case of bilayer materials
such as YBCO and the one layer LSCO.

However, the one layer Tl and Hg cuprates have remained an 
exception\cite{kam} and do not seem to follow the interlayer pair 
tunneling mechanism.  The present author\cite{gb2} made a suggestion
that part of this could be accounted for through the 
particle-hole pair tunneling mechanism.  This still does not
solve the problem completely.  There are also other suggestions
\cite{sudip}. This has become a challenge 
and perhaps a revision of part of Anderson`s Central 
dogma to the effect that `a single layer of cuprate may be 
superconducting' may be called for.  

Thus we have to go back to the original one layer RVB mechanism
and see how it can explain the large $T_c$ of one layer materials.
RVB gauge theory ideas have been pursued a lot along these direction
\cite{patrick}, including some instanton ideas\cite{nagaosa1} and 
an idea\cite{dhl} of quantum tunneling of RVB $\pi$ flux. 
It is becoming clear that an in plane kinetic  mechanism is also
operating in addition to the inter plane kinetic mechanism.
It is also found \cite{gb5} that the inter layer and intralayer 
mechanisms do not help each other and also dominate in different 
regime of doping.

\section{Sharp resonance in Neutron Scattering and quasi particle 
peak in ARPES}

Another outstanding experimental result is the 41 meV resonance.
This sharp 41 meV resonance\cite{kheimer}, limited only by the instrumental
resolution, in neutron scattering in YBCO in the superconducting state 
has a natural explanation
in terms of pair tunneling mechanism as proposed by Anderson
and collaborators.  The peak corresponds to a transition between the 
bonding and antibonding state of an electron  pair between the bilayers,
 induced by the spin flip scattering of the neutron within a layer.  
This resonance is strongly pinned to $(\pi, \pi)$ - this point has no 
satisfactory explanation so far in my opinion.  

Similarly, in the superconducting state, one sees\cite{20mev} a rather 
sharp Bogoliubov quasi particle around $(\pi, 0)$ at a finite
energy of about 20 meV. These quasi particles are rather heavy 
and do not disperse in k-space,  A satisfactory explanation for
this phenomenon, including why this Bogoliubov quasi particle peak 
is confined to the Brillouin zone boundary, away from
normal state fermi surface is not available so far.

\section{Symmetry of the gap function and magnetic field effects}

As it has been emphasized by Anderson and coworkers, the symmetry
of the superconducting order parameter is not strongly dictated by 
the kinetic pairing mechanism.  It view of its strong 
locality in k-space this mechanism determines only the magnitude
of the gap.  The detailed symmetry of the gap function is determined
by the in plane short range repulsion effects, favoring a d-wave. 
That is the local on site constraint $ n_{i\uparrow} + n_{i\downarrow} 
\neq 2 $ leads to a global constraint on the pair amplitude in 
k-space:
$$
<c^\dagger_{i\uparrow} c^\dagger_{i\downarrow}> = 0 \rightarrow 
\sum_k <c^\dagger_{k\uparrow} c^\dagger_{-k\downarrow}> = 0
$$
The above global constraint in k-space is easily satisfied if
the pair amplitude has a d-symmetry.

In the case of conventional s-wave superconductors, small magnetic 
field does not modify the symmetry of the gap function or 
does not collapse the gap.  In the case of the cuprate with 
d-node even small magnetic field strongly modifies the nature
of the superconducting state, as recently discovered by 
Krishana and Ong\cite{krishana} in YBCO.  Their observation of
a magnetic field induced removal of the d-node has
Krishana and Ong\cite{krishana} in YBCO has  revived the 
possibility of $ d_{x^2 - y^2} + i d_{xy}$ state at low temperatures 
and low magnetic fields. Laughlin, Wilczek and others proposal 
of anyonic superconducting state with spontaneous P and T violation
is perhaps realized now, however,  with a small help from a magnetic 
field.  The physics at Dirac nodes of a d-wave superconductor
seems to be filled with rich possibilities\cite{tvr}.

\section{Other approaches to study the t-J model - spin fluctuation, 
 spin bag, SO(5) symmetry etc.}

The spin fluctuation theories\cite{spinfl} work on a fermi liquid 
basis and
suggest pairing mediated by the exchange of spin fluctuation quanta. 
There are deep issues\cite{pwasf} related to the incomplatibility of real space 
super exchange and the fermi liquid background apart from the fact 
that normal state anomalies and inter layer regularities in Tc are 
not explained satisfactorily.  Spin bag theory\cite{spinbag}
is essentially a 
spin fluctuation theory with a real space tinge to it. It also 
suffers from similar criticism.

In the SO(5) front Zhang and collaborators\cite{zhang} view the zero 
temperature superconducting order of the doped 2d cuprates 
as one that is obtained by a rotation of the antiferromagnetic
order.  An alleged $\pi$ operator is the generator of this
rotation.  Serious cricisms\cite{gbso5} starting from technical points 
to points of fundamental principles have been raised.

\section{Conclusion}

While one thought that the mist is getting cleared, one sees some 
further mist that challenges us in the cuprate game. However, the 
direction provided by RVB related ideas has been a constant source
of real understanding of these systems.  And with some more
concerted effort a satisfactory picture should emerge soon.

\section{Acknowledgement}

I thank P.W. Anderson for a critical reading of the manuscript
and comments.

\end{document}